\documentclass{article}

\PassOptionsToPackage{numbers, compress}{natbib}

\usepackage{url}

\usepackage[main, final]{iaseai26}

\usepackage[utf8]{inputenc} 
\usepackage[T1]{fontenc}    
\usepackage{hyperref}       
\usepackage{url}            
\usepackage{booktabs}       
\usepackage{amsfonts}       
\usepackage{nicefrac}       
\usepackage{microtype}      
\usepackage{xcolor}         

\title{Brokerage in the Black Box: Swing States, Strategic Ambiguity, and the Global Politics of AI Governance}

\author{%
  Ha-Chi Tran\\
  The London School of Economics and Political Science\\
  London, United Kingdom\\
  \texttt{c.h.tran@lse.ac.uk} \\
}

\begin{document}

\maketitle

\begin{abstract}
The United States – China rivalry has placed frontier dual-use technologies, particularly Artificial Intelligence (AI), at the center of global power dynamics, as techno-nationalism, supply chain securitization, and competing standards deepen bifurcation within a weaponized interdependence that blurs civilian–military boundaries. Existing research, yet, mostly emphasizes superpower strategies and often overlooks the role of middle powers as crucial actors shaping the global techno-order. This study examines Technological Swing States (TSS) - middle powers with both technological capacity and strategic flexibility - and their ability to navigate the frontier technologies' uncertainty and opacity to mediate great-power techno-competition regionally and globally. It reconceptualizes AI opacity not merely as a technical deficit, but as a structural feature and strategic resource, stemming from algorithmic complexity, political incentives that prioritize performance over explainability, and the limits of post-hoc interpretability. This structural opacity shifts authority from technical demands for explainability to institutional mechanisms - such as certification, auditing, and disclosure - converting technical constraints into strategic political opportunities. Drawing on case studies of South Korea, Singapore, and India, the paper theorizes how TSS exploit the interplay between opacity and institutional transparency through three strategies: (i) delay and hedging, (ii) selective alignment, and (iii) normative intermediation. These practices enable TSS to preserve strategic flexibility, build trust among diverse stakeholders, and broker convergence across competing governance regimes, thereby influencing institutional design, interstate bargaining, and policy outcomes in global AI governance.
\end{abstract}

\section{Introduction}
No longer confined to science fiction, artificial intelligence (AI) has evolved over recent decades from a laboratory innovation into a central axis of twenty-first-century geopolitics. Major powers increasingly treat AI as a strategic asset, linking its development not only to economic competitiveness and security architectures \cite{r:1}, but also to normative influence \cite{r:2}. Unlike earlier episodes of global technology governance shaped by broad multistakeholder cooperation, such as the Montreal Protocol, AI governance presents unprecedented challenges. These stem from AI’s opacity, diffusion, intangibility, rapid innovation cycles, cross-sectoral applications, and the dominant role of private actors \cite{r:3, r:4}. Combined with divergent national security priorities and an increasing narrative of an AI arms race, these features have produced a fragmented, multilayered, and under-representative governance landscape marked by overlapping institutional logics and competing normative agendas, particularly within the context of the United States (U.S.) - China rivalry \cite{r:3}.

This fragmentation is most visible across three overlapping yet divergent paradigms: (i) the U.S.’ market–innovation model, led by technology firms and venture capital but increasingly securitized through export controls and investment screening \cite{r:5, r:6}; (ii) China’s state-led model, which embeds AI in surveillance, defense, and industrial strategies while preserving selective space for private innovation \cite{r:6}; and (iii) the European Union (EU)’s rights-centric model, emphasizing risk-based regulation, fundamental rights protection, and digital sovereignty alongside industrial ambitions \cite{r:7}. These approaches are reinforced by securitization discourses, ranging from existential risk narratives in Western contexts to national security frameworks in the Asia-Pacific, which further impede the emergence of a shared normative foundation.

This technological fragmentation unfolds against the broader backdrop of U.S.–China strategic rivalry, frequently described as a “second Cold War” \cite{r:8, r:9, r:10}. Within this bifurcated landscape, states ostensibly face three strategic options: alignment with the U.S.-led alliance system, integration into a China-centered bloc, or non-alignment. Yet, in practice, sustained non-alignment has become increasingly difficult. Unlike the relatively insulated geopolitics of the original Cold War, contemporary states are deeply embedded in dense global networks of trade, technology, and finance \cite{r:11}. U.S.–China competition is not totalizing; rather, it combines rivalry with coexistence and mutual dependence, as both powers seek to shape the emerging techno-geopolitical order while constraining the strategic space available to smaller states \cite{r:13, r:14}.

Given this context, this paper examines technological swing states as vital intermediaries within the contemporary techno-order and advances a theory of structural brokerage to explain how they convert technological opacity into governance influence under U.S.–China competition. First, it demonstrates that opacity in contemporary AI systems is a durable structural feature rather than a transient technical limitation. This opacity drives a governance shift from model-level transparency to institutional and procedural mechanisms, such as certification schemes, third-party audits, and compliance frameworks, reallocating authority from technical insiders to actors who control verification institutions. Second, the paper theorizes how this situation creates asymmetric opportunities for structural brokerage. When great powers cannot credibly verify one another’s AI capabilities or commitments, technologically capable middle powers, or technological swing states, in advantageous network positions can leverage institutional credibility to mediate, certify, and translate across fragmented governance regimes. A structured comparison of South Korea, Singapore, and India identifies three mechanisms, in particular, delay and hedging, selective alignment, and normative intermediation, through which positional advantage is converted into tangible governance influence.

\section{Opacity by Design: Complexity, Political-economic Incentives, and Limits of Explainable AI}

\subsection{Algorithmic Complexity and the Limits of Explainable AI}
The opacity of contemporary machine-learning systems, particularly large-scale deep learning architectures, should be understood as a structural and enduring property rather than a transient technical shortcoming. As models expand in parameters, training data, and optimization complexity, the input-output mapping becomes increasingly non-linear, distributed, and irreducible to simple causal narratives, producing the so-called “black box” models \cite{r:15, r:16, r:17}. State-of-the-art language and multimodal models now operate with hundreds of billions or trillions of parameters trained on vast, heterogeneous datasets that cannot be exhaustively curated or audited \cite{r:18}. Optimization across high-dimensional spaces with dense cross-layer interactions produces emergent behaviors not explicitly encoded or anticipated \cite{r:20}. Even full access to model weights or architectures may not reveal the causal pathways underlying specific outputs. Opacity is therefore intrinsic to the mathematical and representational complexity of these systems, not merely a product of secrecy or inadequate documentation.

This structural opacity is compounded by the probabilistic nature of modern machine learning. Unlike deterministic, rule-based systems, contemporary sub-symbolic models approximate probability distributions and generate outputs through sampling. Predictions are non-reproducible, and explanations cannot reliably identify causal mechanisms. Outputs reflect dispersed statistical correlations rather than singular causes, creating a persistent mismatch between human expectations of coherent causal reasoning and the statistical logic of model inference \cite{r:21}. Empirical studies illustrate this divergence: humans classify animals using semantic features, such as eyes or fur, whereas neural networks often rely on imperceptible pixel patterns or latent frequency cues. For example, models have misclassified huskies as wolves when images contained snow in the background, demonstrating reliance on spurious correlations rather than salient features \cite{r:22}. Even when influential features are identified, they rarely yield cognitively meaningful explanations, as these signals do not map onto human categories of perception or reasoning.

Explainable AI (XAI) was initially heralded as a remedy for such opacity \cite{r:25, r:26, r:27}. However, contemporary XAI techniques are largely post hoc: they generate explanations after the fact to rationalize outputs rather than reconstructing a model’s internal reasoning \cite{r:26}. Consequently, XAI outputs often serve less as disclosures of epistemic processes than as plausible rationalizations calibrated to human expectations of causality. Current XAI strategies cluster around two approaches. First, model simplification approximates complex architectures with interpretable surrogates, such as linear regressions, decision trees, or locally faithful proxies \cite{r:29, r:25}. While cognitively tractable, these proxies inevitably sacrifice fidelity by omitting higher-order interactions and latent features, producing an “ersatz understanding” of black-box models \cite{r:25}. Second, feature attribution methods, such as SHAP and LIME, assign marginal contributions to inputs \cite{r:29, r:33, r:22}. Although mathematically rigorous, they highlight correlations rather than causal mechanisms and rarely answer the “why” questions critical for legal, ethical, and policy reasoning. Empirical studies also show that different XAI methods can yield contradictory feature attributions for the same model outputs \cite{r:92}, raising doubts about whether explanations reflect true decision processes or merely post hoc rationalizations.

This tension between interpretability and fidelity, indeed, has governance implications. Explanations optimized for human understanding often obscure true model mechanisms, while faithful accounts remain cognitively inaccessible. Post hoc explanations deployed for compliance are legally fragile and ethically precarious, leaving systems vulnerable to manipulation, bias concealment, and adversarial exploitation. High-profile cases, including the COMPAS risk-assessment tool in U.S. criminal justice and proprietary credit-scoring platforms, illustrate how interpretability features can legitimize opaque systems under the guise of transparency, entrenching systemic inequalities \cite{r:26}.

\subsection{Political-Economic Logic of Opacity: Performance Over Transparency}

The persistence of AI opacity cannot be explained by technical complexity alone. Political–economic dynamics play an equally decisive role. Rather than a neutral byproduct of engineering constraints, opacity reflects structural incentives that prioritize speed, scalability, and competitive advantage over interpretability. In a global AI economy structured around first-mover advantages and rapid iteration, transparency is increasingly framed not as a public good but as a source of friction. These incentives are further amplified by the securitization of AI, which positions it simultaneously as critical infrastructure and as a site of geopolitical rivalry. Under such conditions, opacity no longer signals a governance failure; it becomes an institutionalized feature, deliberately cultivated through state policy, corporate practice, and techno-industrial competition.

A central tension concerns whether interpretability and performance can be jointly optimized. While contested, a prevailing view holds that interpretability often entails trade-offs with efficiency and accuracy, the trade-offs embedded in the economic imperatives shaping AI development \cite{r:34, r:35}. Venture-capital-driven commercialization privileges rapid scaling and market dominance, reinforcing incentives to favor capability over explainability. Advocates of inherently interpretable or “white-box” AI argue that, in high-stakes domains such as healthcare or criminal justice, sacrificing transparency for marginal performance gains is normatively indefensible \cite{r:26}. Empirical evidence, however, suggests that economic and competitive pressures routinely override these concerns. In autonomous driving, for example, Tesla relies on opaque end-to-end neural architectures rather than modular or rule-based alternatives \cite{r:38}, despite recurring safety controversies \cite{r:71, r:72}.

Geopolitical competition further entrenches these dynamics. As AI is increasingly treated as dual-use infrastructure, disclosure of model architectures, training data, or parameter weights is reframed as a security vulnerability. In the U.S., Executive Order 14110 (2023) mandates federal agencies to assess and report risks associated with frontier AI models, exemplifying oversight without full public transparency. Recommendations from the U.S. National Telecommunications and Information Administration similarly endorse restricting access to certain model weights \textit{``if the U.S. government assesses that the risks of their wide availability sufficiently outweigh the benefits,''} through distribution controls, export licensing, corporate access limits, and regulated APIs \cite{r:41}. At the 2025 Paris AI Action Summit, Vice President J.D. Vance echoed this approach, emphasizing guardrails while avoiding commitments to algorithmic transparency, reflecting a strategy that prioritizes innovation alongside national security and economic competitiveness \cite{r:42}.

Even within the EU—a jurisdiction traditionally associated with strong norms of transparency and accountability—opacity persists where industrial competitiveness and strategic positioning are implicated. Early drafts of the AI Act included stringent interpretability and explainability requirements for high-risk systems, reflecting normative commitments to public oversight and safety \cite{r:43}. Yet these provisions risk becoming largely symbolic. Article 78 of the final AI Act permits disclosure only when deemed “strictly necessary,” \cite{r:93} a standard left undefined in the legislation. Indeed, firms can invoke trade-secret protections with minimal evidentiary burden, allowing assertions of confidentiality to block access or delay enforcement \cite{r:94}.

These compromises reflect broader economic and geopolitical pressures. European policymakers and firms expressed concern that rigid disclosure obligations could undermine competitiveness or impose chilling effects on smaller companies \cite{r:48, r:49}, particularly relative to faster-moving U.S. and Chinese competitors. A 2022 joint survey by Applied AI found that half of AI startups believed the AI Act would slow innovation in the EU, while 16\% reported considering halting AI development or relocating operations abroad \cite{r:46}. That is to say, even in a jurisdiction with strong formal commitments to transparency, regulatory discourse has struggled to reconcile normative aspirations with market and geopolitical imperatives favoring opacity.

Corporate narratives increasingly align with state logics by reframing opacity in generative AI not as a limitation but as a necessary safeguard. This discursive shift legitimizes the enclosure of foundational technologies under the dual banner of public safety and responsible innovation. In practice, however, opacity functions as a competitive asset. Open-source large language models consistently lag proprietary systems by five to twelve months in benchmark performance \cite{r:39}, underscoring the market value of restricted access. Leading firms actively shape this logic through selective disclosure. OpenAI frames the limited release of GPT-4 as a safety imperative \cite{r:39, r:40}. Similarly, Google, Meta, and Anthropic engage in partial disclosures that fall short of open-source principles, withholding critical components such as architectures and codebases while imposing licensing regimes that restrict modification, redistribution, and commercial use \cite{r:50, r:51, r:52}.

These practices are justified not only on commercial grounds but also through concerns about misuse, including the risk that access to model weights could enable adversarial fine-tuning or the circumvention of safeguards. Firms further argue that releasing the source code of one system may expose vulnerabilities in others \cite{r:51}. Together, these narratives normalize algorithmic secrecy as both commercially rational and ethically defensible, reinforcing a governance regime in which privatized control over core AI infrastructure is rendered not merely acceptable, but increasingly inevitable.

\subsection{From Model Transparency and Explainability to Institutional and Procedural Oversight}
When technical, economic, and geopolitical pressures transform opacity from a contingent design feature into a legitimized norm, it becomes an asset to be managed rather than a deficit to be eliminated. Regulatory strategies premised on universal explainability therefore encounter systemic resistance, as speed, scale, and capability remain existential imperatives. The central governance challenge is not the eradication of opacity, but the design of oversight mechanisms that function under conditions in which opacity is structurally incentivized and politically valorized. In this context, institutional and procedural transparency becomes decisive. Where model-level transparency is technically unattainable, legitimacy is secured through formal arrangements that render decisions auditable and contestable—by documenting design choices, clarifying responsibility, and establishing mechanisms for oversight, appeal, and redress. Under such conditions, accountability derives less from understanding internal model logic than from the integrity, rigor, and auditability of governance processes.

Pharmaceutical regulation provides a useful analogy in which regulators did not require full causal clarity \cite{r:28}. Instead, they constructed layered governance regimes—clinical trials, phased approvals, disclosure obligations, post-market surveillance, and compensation frameworks—that rendered decisions transparent and contestable despite epistemic uncertainty. These procedural systems were designed to adapt continuously as evidence evolved. Legitimacy rested not on causal comprehension but on institutional capacity to manage uncertainty responsibly. Applied to AI, this logic treats opaque models as complex scientific systems whose deployment is governed through rigorous research practices and methodological safeguards rather than full interpretability. Procedural mechanisms thus function as compensatory structures, generating trust grounded in empirical adequacy, predictive reliability, and institutional coherence rather than internal transparency.

This shift toward procedural transparency is evident across global AI governance regimes. In the U.S., Executive Order 14110 operationalizes accountability through standardized testing, evaluation, provenance tracking, and inter-agency reporting, rather than disclosure of model weights or training data \cite{r:73}. Practices such as red-teaming, incident reporting, and model labeling reflect a broader turn toward process-based oversight. In China, accountability is enforced through centralized monitoring, algorithm filing, and state-auditable security assessments under the Cyberspace Administration of China \cite{r:75}. The EU’s AI Act similarly emphasizes risk-based documentation, third-party audits, and impact assessments \cite{r:76}. International organizations, including the OECD and UNESCO, advance normative frameworks centered on procedural accountability rather than technical interpretability \cite{r:77}. Across these contexts, adaptive tools—regulatory sandboxes, phased approvals, and experimental oversight—demonstrate that transparency increasingly functions as a property of institutional design, not model explainability.

Nevertheless, these institutional architectures both manage and reproduce opacity. Governance design determines who must disclose what, to whom, under which evidentiary standards, and within which liability regimes. In doing so, it structures the distribution of transparency, risk, and legitimacy—privileging states, firms, or publics; socializing or privatizing harm; and grounding legitimacy in procedural rigor, normative commitments, or both. Transparency is therefore not absolute but strategically conditioned by governance architecture.

Within this environment, technical opacity generates additional layers of uncertainty that shape state behavior. First, informational opacity persists because frontier models remain only partially knowable, even to their developers. Second, normative opacity arises because deep epistemic uncertainty prevents stable agreement on what constitutes safe, fair, or trustworthy AI. When system behavior is probabilistic, path-dependent, and influenced by latent variables, evaluative standards become contingent on political philosophy, regulatory capacity, and risk tolerance. This indeterminacy is reflected in diverging regulatory templates—ex ante versus ex post oversight, risk-based versus capability-based regimes, and centralized versus decentralized governance—each embodying different assumptions about how AI risk emerges and can be contained.

Third, informational and normative uncertainty create the conditions for strategic opacity. States face ambiguity about which technological ecosystems will prevail, which regulatory paradigms will crystallize into global norms, and which geopolitical alignments will secure long-term access to compute, talent, supply chains, and markets. Under export controls, supply-chain securitization, and techno-industrial competition, premature commitments risk high and potentially irreversible costs, including technological lock-in, regulatory incompatibility, and political misalignment. As a result, states may deliberately cultivate ambiguity: delaying commitments to preserve flexibility, maintain access to multiple ecosystems, and avoid costly entanglements. This strategic reticence produces an international environment marked by mutual uncertainty, in which actors remain unsure not only of their own trajectories, but also of the intentions, constraints, and alignments of rivals and partners. Building on this analysis, the following section delves into case studies of South Korea, Singapore, and India, examining how these states navigate opacity and strengthen their techno-power.

\section{Opacity as Resource: Brokerage Power in A Fragmented AI Governance}
\subsection{Network Position and the Rise of Technological Swing States}
If the contemporary international system is conceived as a hyper-connected network anchored by two central yet rivalrous nodes, the U.S. and China, then states situated between them occupy intermediary positions defined by dense but asymmetrical ties to both powers. Unlike earlier eras of bipolarity, where states could shelter within relatively discrete blocs, today’s system forms a single giant component, interwoven through global supply chains, transnational data infrastructures, multilateral institutions, and overlapping diplomatic commitments \cite{r:13, r:95}. It is characterized by high connectivity, short path lengths, and multi-scalar interdependencies: even adversarial powers remain directly linked via trade, standards bodies, financial architectures, and shared technological vulnerabilities \cite{r:96}. Influence thus derives less from hierarchical dominance than from structural position: how actors are embedded in flows of information, resources, authority, and technological governance.

As shown in the preceding section, AI’s structural opacity not only limits model-level explainability but also reshapes global governance. The shift from epistemic to institutional and procedural transparency is inseparable from informational, normative, and strategic opacity, each reinforcing the others. Yet this partially opaque environment is not solely constraining. For technologically capable but geopolitically exposed jurisdictions, layered opacities create levers of influence, enabling strategic maneuvering, standard-setting participation, and selective alignment across competing ecosystems. This context renders the concept of technological swing states analytically salient.

Originating in U.S. domestic politics, “swing states” denote actors whose influence stems from the absence of durable alignment with any single power \cite{r:97}. In international relations, the term describes states engaging major powers without exclusive allegiance, preserving strategic autonomy \cite{r:53, r:98}. This paper adapts the concept to the U.S.–China technological rivalry, defining technological swing states as intermediate actors: they possess substantial technical and political capacity but lack coercive authority to impose system-wide rules. Their leverage derives from flexibility, brokerage, and the ability to navigate contested, opaque governance environments.

Technological swing states combine several attributes, though not always simultaneously, that shape their strategic leverage. These include advanced domestic research infrastructure, innovation ecosystems, and skilled human capital; geostrategic positioning within critical technology supply chains, where they may act as production hubs, component suppliers, or gatekeepers; control of strategic infrastructures such as satellites, submarine cables, or data centers; and influence in governance forums through regulatory debates, technical standards, and multilateral negotiations. The particular configuration of these attributes determines the scope and form of leverage. Occupying this intermediate space, such states exercise substantial but non-coercive power, advancing national technological agendas while retaining autonomy from great-power dominance.

This paper uses network theory analysis as an analytical heuristic rather than a formal modeling framework. It examines how brokerage mechanisms—delay and hedging, selective alignment, and normative intermediation—operate within these structural positions through process tracing and in-depth case studies. This approach aligns with structural positional analysis in international relations, where network concepts provide a vocabulary for reasoning about position without formal quantification \cite{r:100, r:101}. Future research may operationalize these arguments quantitatively, but here the focus is on South Korea, Singapore, and India—technologically capable yet geopolitically exposed states pursuing distinct strategies amid intensifying U.S.–China competition.

\subsection{Delay and Hedging: The Case of South Korea}

States that exploit delay, ambiguity, and hedging tend to occupy strategic positions along key pathways of economic interdependence, technological diffusion, and geopolitical rivalry. Rather than operating as alternative hubs, they mediate great-power competition from locations embedded near the network core of dominant actors. Where these pathways connect intermediary states directly to both the U.S. and China, influence stems less from bridging disconnected peripheries than from proximity to the most powerful nodes in the global technology network.

This embeddedness simultaneously generates leverage and constraint. Such states cannot fully align with one power without incurring significant economic, technological, and security costs, nor can they disengage without sacrificing influence. Their agency derives from mutual dependence with great powers, yet this same dependence exposes them to coercion and limits autonomy. Alignment choices are therefore path-dependent: premature commitments risk irreversible lock-in with far-reaching consequences. Formal analyses \cite{r:104, r:105} demonstrate that alignment with a technological, regulatory, or geopolitical bloc entails high sunk costs—ranging from standards adoption and supply-chain reconfiguration to regulatory harmonization and costly political signaling—thereby constraining future strategic adjustment.

Consequently, intermediary states rationally prioritize preserving option value. By maintaining multiple channels of engagement and deferring irrevocable commitments, they ensure against technological discontinuities, shifts in relative power, and the volatility of great-power rivalry. Strategic ambiguity thus reflects optimization under uncertainty rather than indecision: delaying firm alignment while issuing equivocal signals maximizes expected payoffs and sustains flexibility. At a theoretical level, hedging functions as an insurance mechanism \cite{r:106, r:107}, protecting future strategic options against premature lock-in in fluid technological and geopolitical environments.

South Korea’s posture in the U.S.–China technology rivalry exemplifies this logic. Occupying a pivotal brokerage position in global semiconductor and AI networks, Seoul leverages strategic opacity to convert uncertainty in frontier-technology governance into bargaining power. As competition intensifies, the structural value of such intermediary nodes increases: misalignment becomes more costly for both great powers, reshaping the payoff landscape. Washington and Beijing must therefore internalize the risks of alienating a technologically indispensable actor whose delayed commitments materially affect outcomes in AI hardware, computing capacity, and next-generation telecommunications standards.

South Korea’s leverage in the U.S.–China technology rivalry rests on its industrial and technological depth. As the world’s second-largest semiconductor producer, anchored by Samsung and SK Hynix \cite{r:54}, Korea is projected to account for roughly 19\% of global output by 2025, including over 60\% of the memory segment \cite{r:44}. Its AI “Innovation and Integration”, ranked third globally \cite{r:45}, supported by sustained R\&D investment, extensive infrastructure, and a rapidly scaling startup sector. 

This leverage is reinforced by Korea’s dual dependence on China and the U.S., which renders unilateral alignment both costly and inefficient \cite{r:79}. In 2024, more than half of Korea’s semiconductor exports went to China \cite{r:55}, reflecting deep integration into East Asian manufacturing ecosystems. Simultaneously, Korea remains a core U.S. ally, embedded in Western research networks, security architectures, and export-control regimes. This dual entanglement creates structural irreversibility: decisive alignment with either power risks severe economic or political retaliation from the other, making delay and hedging a rational strategy rather than a temporary expedient.

The strategic value of delay became evident following U.S. chip and equipment export controls introduced in 2022 and tightened in 2023. These measures have negatively affected Korean semiconductor exports to China, particularly high-value memory products \cite{r:83, r:79}, and threatened the viability of Korean fabs operating in China \cite{r:78}. Seoul responded through targeted diplomatic hedging, securing conditional U.S. waivers. Programs such as the Validated End-User (VEU) designation and the Restricted Fabrication Facility (RFF) provision allowed Samsung and SK Hynix to continue supplying U.S. equipment to their Chinese plants without case-by-case licensing \cite{r:80, r:81, r:82}. Though temporary, these exemptions functioned as real options, buying time to absorb shocks, reallocate capital, and sequence multi-billion-dollar supply-chain decisions without premature lock-in.

Delay and hedging operate through a dual logic. Temporally, delay provides a buffer for monitoring U.S.–China relations, assessing the durability of export-control regimes, and managing technological uncertainty before committing irreversibly. Structurally, hedging preserves diversified technological and commercial pathways, reducing vulnerability to coercion. For Washington, excessive tightening risks pushing Seoul toward Beijing, while excessive leniency weakens coalition discipline. For China, pressure risks accelerating Korean compliance with U.S. controls. Korea’s strategy thus reshapes both powers’ payoff structures by raising the costs of coercion and increasing the value of accommodation.

This position, yet, remains fragile. In mid-2025, signals from the Trump administration regarding potential revocation of VEU and RFF exemptions underscored the vulnerability of arrangements dependent on U.S. executive discretion \cite{r:66, r:67}. Seoul responded with deeper diplomatic and industrial hedging. Under the AI Diffusion Framework introduced in early 2025, South Korea secured Tier 1 status as a trusted partner, gaining partial insulation from forthcoming AI export controls \cite{r:84}. In parallel, the Sovereign AI Strategy, including a 100-trillion-won National Growth Fund, aims to develop domestic foundation models through firms such as LG AI Research and Naver Cloud \cite{r:85, r:86}. The AI Framework Act (January 2025) further strengthens autonomy by establishing independent risk-management obligations for high-impact AI systems, enhancing Korea’s regulatory credibility in multilateral governance forums \cite{r:87}.

\subsection{Selective Alignment: The Case of Singapore}
Selective alignment operates through three mutually reinforcing mechanisms: active multi-homing, task-specific partnerships, and modular network positioning. Active multi-homing enables engagement with multiple partners across economic, technological, and security domains, reducing dependence on any single power and vulnerability to coercion. Task-specific partnerships facilitate pragmatic, issue-focused cooperation—such as AI governance, digital trade, or cybersecurity—without entailing broader ideological alignment. Modular network positioning amplifies agency by situating states as connectors within regional and global governance networks, allowing them to coordinate fragmented actors, translate regional concerns into global norm-setting, and diffuse international standards back into local contexts.

Unlike South Korea’s delay-and-hedging strategy, which reflects structural exposure to dual U.S.–China dependencies, Singapore adopts a proactive, engagement-oriented approach, exemplifying selective alignment. This strategy enables cooperation with competing powers while preserving autonomy and shaping governance architectures without entering irreversible alignment paths. This flexibility is enabled by the nature of Singapore’s technological power. Whereas South Korea’s hardware-based, capital-intensive industries are deeply embedded in global supply chains and characterized by high switching costs, Singapore’s strengths—digital infrastructure, data governance, cybersecurity, and AI standards—are modular, software-driven, and interoperable. As a result, Singapore faces lower sovereignty costs, greater reversibility in partnerships, and reduced exposure to strategic entrapment. Its technological power derives from infrastructural centrality and governance credibility rather than fixed manufacturing dependencies, expanding its strategic margin of maneuver.

Singapore’s structural power is grounded in advanced physical and institutional infrastructure. Its extensive submarine cable network positions the city-state as a central node in global data flows, reinforced by a dense ecosystem of hyperscale data centers operated by AWS, Google Cloud, and Microsoft Azure \cite{r:57, r:58, r:59}. These foundations support regional cloud services, AI model training, and large-scale data storage. Singapore’s research ecosystem further strengthens its position: it ranks among the global top five in novel metamaterials \cite{r:108}, with Nanyang Technological University and National University of Singapore recognized as world-leading institutions, and has consistently placed in the top ten of the AI Vibrancy Index since 2019 to 2023 \cite{r:109}.

Selective alignment is reinforced by Singapore’s dual role as a central ASEAN node and a credible interlocutor in global digital governance. Through multi-homing, it engages major powers across domains; through task-specific partnerships, it collaborates selectively with the U.S., China, and the EU without broader ideological commitments. Its infrastructural embeddedness generates material interdependencies that regional regulators, multinational firms, and neighboring states cannot bypass. Singapore has assumed leadership roles in global forums such as the Bletchley Park Summit despite limited participation from other Southeast Asian states \cite{r:60}. As AI systems remain opaque, partners increasingly rely on institutional mechanisms, certification, auditing, and compliance, enhancing the value of Singapore’s procedural architectures as sources of normative authority.

Institutional transparency operationalizes selective alignment by offering auditable governance without intrusive technical disclosure. Certification schemes, independent audits, and structured risk-management protocols signal reliability across regimes while enabling Singapore to bridge OECD ethical principles, ISO technical standards, and ASEAN data-protection rules within a coherent framework. By positioning institutional transparency as a functional substitute for unattainable model-level disclosure, Singapore converts opacity into a bargaining resource and renders its governance frameworks indispensable in a fragmented regulatory environment.

The proactive nature of selective alignment distinguishes Singapore from defensive hedging strategies. Rather than deferring commitments, Singapore aligns incrementally across blocs to shape outcomes. The Model AI Governance Framework, introduced in 2019 and iteratively updated, integrates global best practices while adapting them to local conditions and has been adopted by industry consortia as a governance template \cite{r:61}. Singapore’s leadership within the World Economic Forum and other forums further embeds its institutional designs within transnational governance infrastructures, positioning it as a rule-shaper rather than a rule-taker \cite{r:62, r:63}.

The Smart Nation initiative also exemplifies the operationalization of selective alignment. By integrating AI and digital technologies across public services, urban management, and data governance, Singapore institutionalizes procedural transparency and regulatory oversight while leveraging proprietary innovation \cite{r:63}. Additionally, infrastructural centrality magnifies the effects of selective alignment. Because data flows, cloud services, and AI deployments across Southeast Asia frequently transit through Singapore’s networks, domestic certifications and audit protocols acquire regional authority \cite{r:57, r:58}. By embedding accountability into these infrastructures, Singapore exports its governance preferences and reinforces normative influence. In this context, AI opacity becomes a source of leverage: actors rely on Singapore’s institutional credibility precisely because technical internals cannot be independently verified.

Selective alignment delivers two core strategic effects. It preserves autonomy by preventing overdependence on any single bloc while ensuring interoperability across regimes, and it enhances regional influence by positioning Singapore as a mediator of standards within ASEAN and between Asia and the transnational governance space. By functioning as a platform for interoperability and a trusted provider of institutional transparency, Singapore converts uncertainty into durable authority. Unlike passive delay strategies, selective alignment actively leverages opacity to negotiate and shape outcomes, demonstrating how small but technologically capable states can transform structural asymmetries in the global AI order into sustained agency and influence.

\subsection{Normative Intermediation: The Case of India}
India’s approach to AI governance exemplifies normative intermediation, a strategy that operates not through unilateral rule export or tight policy convergence, but through translation, discourse shaping, and multilateral brokerage. Whereas Singapore’s selective alignment derives influence from interoperable rule design, procedural transparency, and modular regulatory adoption, India intervenes at a different level of the governance game. It leverages deep embeddedness across diverse diplomatic and epistemic networks to shape focal points, interpretive frames, and payoff structures under technological uncertainty. If Singapore mitigates technical opacity through auditability, India addresses normative opacity—the uncertainty surrounding which principles should govern, which objectives matter, and how competing models should be reconciled. Selective alignment stabilizes expectations around existing standards; normative intermediation creates, translates, and reweights standards themselves.

This divergence reflects distinct structural profiles. Singapore’s authority stems from infrastructural centrality as a logistics, data, and regulatory hub embedded in dense regional supply chains, enabling it to export trusted procedures as compatibility layers. India, by contrast, exhibits multi-network embeddedness: it is simultaneously a central actor in Global South diplomacy, an increasingly indispensable partner in U.S.-led technology coalitions such as the Quad and iCET, a participant in ASEAN-centered digital forums, and a leader in multilateral platforms like the G20. Its influence does not derive from controlling a single chokepoint but from bridging heterogeneous governance communities with divergent normative expectations. This position enables normative entrepreneurship, supported by large epistemic communities in AI talent, global diaspora networks, and digital public infrastructure expertise.

Normative intermediation functions through preference transformation rather than procedural harmonization. India alters governance outcomes by reshaping how actors define the problem itself. In contexts where AI systems, algorithmic risks, and cross-border data flows remain conceptually opaque, India proposes focal normative frames, such as “AI for All” \cite{r:110} or “AI for Development,” \cite{r:111} that reduce interpretive ambiguity and reweight the payoff matrix. Once such frames become focal, powerful actors must incorporate them to maintain legitimacy, effectively transforming ambiguity into leverage by defining what counts as an acceptable governance objective.

The 2023 G20 New Delhi Summit illustrates this mechanism \cite{r:89}. India used its agenda-setting authority to shift discourse away from purely precautionary or security-centric frames toward development, inclusion, and responsible innovation \cite{r:65, r:88}. This reframing did not merely mediate preferences but altered them: development-sensitive language became embedded in subsequent global AI negotiations, including those on data governance and risk assessment. India’s planned 2026 Global AI Summit is poised to extend this momentum, reinforcing its brokerage role.

India’s domestic governance architecture further strengthens normative intermediation by providing procedural authority without full algorithmic disclosure. Through the National AI Strategy, sector-specific ethical guidelines, and hybrid public–private audit mechanisms, India offers credible frameworks for risk assessment, certification, and compliance translation \cite{r:65}. These function as interpretive infrastructure, making India’s reading of opaque systems indispensable to cross-border cooperation. As network theory predicts, India increases centrality not by controlling information, but by controlling the interface through which norms are interpreted.

This leverage is evident in India’s role within the Quad AI Working Group. While collaborating with the U.S., Japan, and Australia on trustworthy AI principles \cite{r:90}, India adapts these norms for broader Indo-Pacific application \cite{r:91}, including regions with differing regulatory capacities and development priorities \cite{r:64}. Similar dynamics are visible in ASEAN engagement, where India advances “ethical interoperability,” blending GDPR-style protections, OECD principles, and local innovation needs to reduce normative friction without requiring strict alignment.

India’s technological profile reinforces this role. Although not dominant in advanced hardware or frontier foundation models, India is a global leader in software services, AI engineering talent, and digital public infrastructure. These capabilities generate epistemic legitimacy, enhancing its authority to mediate technical and normative issues. At the same time, closer alignment with the U.S. through initiatives such as iCET, CHIPS-linked semiconductor cooperation, and INDUS-X strengthens India’s position within Western technology networks, while its continued leadership in the Global South ensures broad receptivity to its normative proposals. This dual embeddedness allows alignment without loss of autonomy.

The contrast with Singapore is therefore clear. Selective alignment is an operationally precise strategy centered on harmonization, accountability, and procedural credibility, scaling outward through adoption. Normative intermediation is a political-discursive strategy aimed at redefining what governance is about. Its authority derives not from infrastructural centrality but from the ability to reduce normative opacity, propose focal frames, and bridge ideational divides across blocs. Through this strategy, India converts fragmentation—divergent regulations, opaque technologies, heterogeneous preferences—into a source of power, exerting influence disproportionate to its material capabilities and positioning itself as a central node in the emerging architecture of global AI governance.

\section{Conclusion and Limitation}
This paper examines how technological swing states exercise strategic agency within an increasingly fragmented and polycentric global AI governance landscape. It argues that such states convert structural opacity into strategic leverage and use institutional transparency as a functional substitute for otherwise unavailable technical disclosure, thereby transforming uncertainty into procedural and positional power.

The analysis does not assume that the outcomes observed in the three case-study countries are mutually exclusive, nor does it presuppose that they reflect fully articulated, ex ante strategies or coherent self-understandings of agency. Rather, it acknowledges that the mechanisms at play may operate simultaneously or evolve over time. The study identifies the underlying mechanisms through which influence is exerted, irrespective of deliberate intent, as states navigate the constraints and opportunities inherent in their structural positions. In this context, influence emerges from the interplay between technological opacity and institutional transparency. Importantly, the argument does not assert a direct, linear relationship between network centrality and the exercise of leverage, nor does it assume that strategic behavior is the result of conscious, intentional choice. Instead, the analysis examines how structural positions, institutional practices, and observable governance outcomes align in shaping influence. This approach refrains from advancing a fully specified causal theory of strategic behavior, focusing instead on understanding the complex, contingent factors that contribute to the exercise of power within a fragmented global governance landscape.

However, several limitations remain. The paper focuses on three case studies, offering valuable comparative insights, but it remains selective and provides only an initial exploration rather than a comprehensive analysis of each state's strategic environment. By centering on sovereign states, the paper underemphasizes the role of supranational and interstate actors, most notably the EU, whose collective regulatory capacities may shape or constrain the maneuvering space of individual swing states. This state-centric focus also excludes other emerging technological actors, such as Canada, Brazil, Indonesia, and the United Arab Emirates, whose trajectories may diverge due to distinct regional institutions, political economies, or technological endowments. Furthermore, the analysis gives limited attention to nonstate actors, including multinational firms, transnational standard-setting bodies, and civil society coalitions, which increasingly influence the operational and normative boundaries of global AI governance.

Future research should extend both the empirical and theoretical reach of this inquiry. Empirically, examining additional cases and regional contexts would enable deeper process tracing of how intermediary positions are constructed, sustained, or eroded over time, and how varying network configurations condition swing-state influence. Theoretically, this study should be read as a contribution to case-based analysis and semi-formalization rather than a fully developed network theory. Concepts such as centrality, opacity, and intermediation function as analytical lenses rather than foundational frameworks. Further work is needed to refine these concepts into testable propositions, specify causal mechanisms more explicitly, and integrate insights from transnational governance, political economy, and science and technology studies.

\bibliographystyle{plainnat}
\bibliography{Tran_final}

\end{document}